\documentclass{article}

\usepackage{arxiv}

\usepackage{amssymb}

\usepackage{array}
\usepackage[binary-units]{siunitx}
\DeclareSIUnit[number-unit-product = {}]{\inchQ}{''}
\usepackage[nolist]{acronym}
\usepackage{hyperref}
\usepackage{makecell}
\usepackage{graphicx}
\graphicspath{{./Figures/}}

\title{Finite state machine controls for a source of optical squeezed vacuum}

\author{
 Mateusz~Bawaj \\
  INFN, Sezione di Perugia \\
  \& Universit\`a di Perugia \\
  I-06123 Perugia, Italy \\
  \texttt{mateusz.bawaj@pg.infn.it} \\
   \And
 Valeria~Sequino \\
  INFN, Sezione di Napoli, \\
  \& Universit\`a di Napoli Federico II \\
  Complesso Universitario di Monte S.Angelo \\
  I-80126 Napoli, Italy \\
  \texttt{valeria.sequino@na.infn.it} \\
   \And
 Catherine~Nguyen \\
  Universit\'e de Paris, CNRS \\
  AstroParticule et Cosmologie \\
  F-75013 Paris, France \\
   \And
 Marco~Vardaro \\
  Nikhef, Amsterdam, \\
  \& Institute for High-Energy Physics \\
  University of Amsterdam \\
  Amsterdam, The Netherlands \\
   \And
 Sibilla~Di Pace \\
  INFN, Sezione di Roma \\
  \& Universit\`a di Roma La Sapienza \\
  I-00185 Roma, Italy\\
   \And
 Imran~Khan \\
  Gran Sasso Science Institute (GSSI) \\
  I-67100 L’Aquila \\
  \& INFN, Sezione di Roma Tor~Vergata \\
  I-00133 Roma, Italy \\
   \And
 Diego~Passuello \\
  INFN, Sezione di Pisa \\
  I-56127 Pisa, Italy \\
   \And
 Alberto~Gennai \\
  INFN, Sezione di Pisa \\
  I-56127 Pisa, Italy \\
   \And
 Martina~De Laurentis \\
  INFN, Sezione di Napoli, \\
  \& Universit\`a di Napoli Federico II \\
  Complesso Universitario di Monte S.Angelo \\
  I-80126 Napoli, Italy \\
   \And
 Jean-Pierre~Zendri \\
  INFN, Sezione di Padova \\
  I-35131 Padova, Italy \\
   \And
 Fiodor~Sorrentino \\
  INFN Sezione di Genova \\
  I-16146 Genova, Italy\\
}

\begin{document}
\maketitle

\date{\today}

\begin{abstract}
\noindent 
In this paper we present a software, developed in the distributed control system environment of the Virgo gravitational-wave detector, for the management of a highly automated optical bench. The bench is extensively used for the research and development of squeezed states of light generation in order to mitigate the quantum noise in the next generations of interferometric gravitational-wave detectors. The software is developed using \aclp{FSM}, recently implemented as a new feature of \textit{damping-adv} \acl{SDK}. It has been studied for its ease of use and stability of operation and thus offers a high duty-cycle of operation. Much attention has been drawn to ensure the software scalability and integration with the existing \acl{DAQ} and control infrastructure of the Virgo detector.
\end{abstract}



\section{Motivation and significance}
\label{sec:intro}
\noindent
Squeezed states of light, in audio frequency band, using an \ac{OPO}, were realized for the first time in 2004~\cite{mckenzie2004squeezing}. With improvements in experimental techniques, that allowed to expand the squeezing bandwidth~\cite{vahlbruch2006coherent,vahlbruch2007quantum}, squeezed states of light are nowadays used to reduce \ac{QN} in the \ac{GW} detectors network~\cite{GEO600SqzInj,acernese2019increasing,tse2019quantum}; hence raising interests for further research and development in this field~\cite{Valeria,Sibilla}. Towards that aim, a dedicated experimental test facility is developed at the European Gravitational Observatory hosting the Virgo \ac{GW} detector. The effective use of such a source of light requires an easy-to-use and reliable control apparatus. Therefore, in this paper, we report a software, that is developed for a fully automated squeezing facility. The control design is focused on making the setup as accessible as possible to scientists from various fields of expertise, working on the experimental setup. Thus, the software implementation reduces any complex operation on the feedback controls, such as tuning or calibration. Automation minimizes the human interaction with the locking during operators activities on the bench and helps in optimization of the subsystems' \ac{QoS}. We describe the experiment in the following section.

\section{Experimental setup}
\label{sec:apparatus}
\subsection{Optical setup}
\begin{figure}
  \begin{minipage}[t]{0.67\textwidth}
    \includegraphics[width=\textwidth]{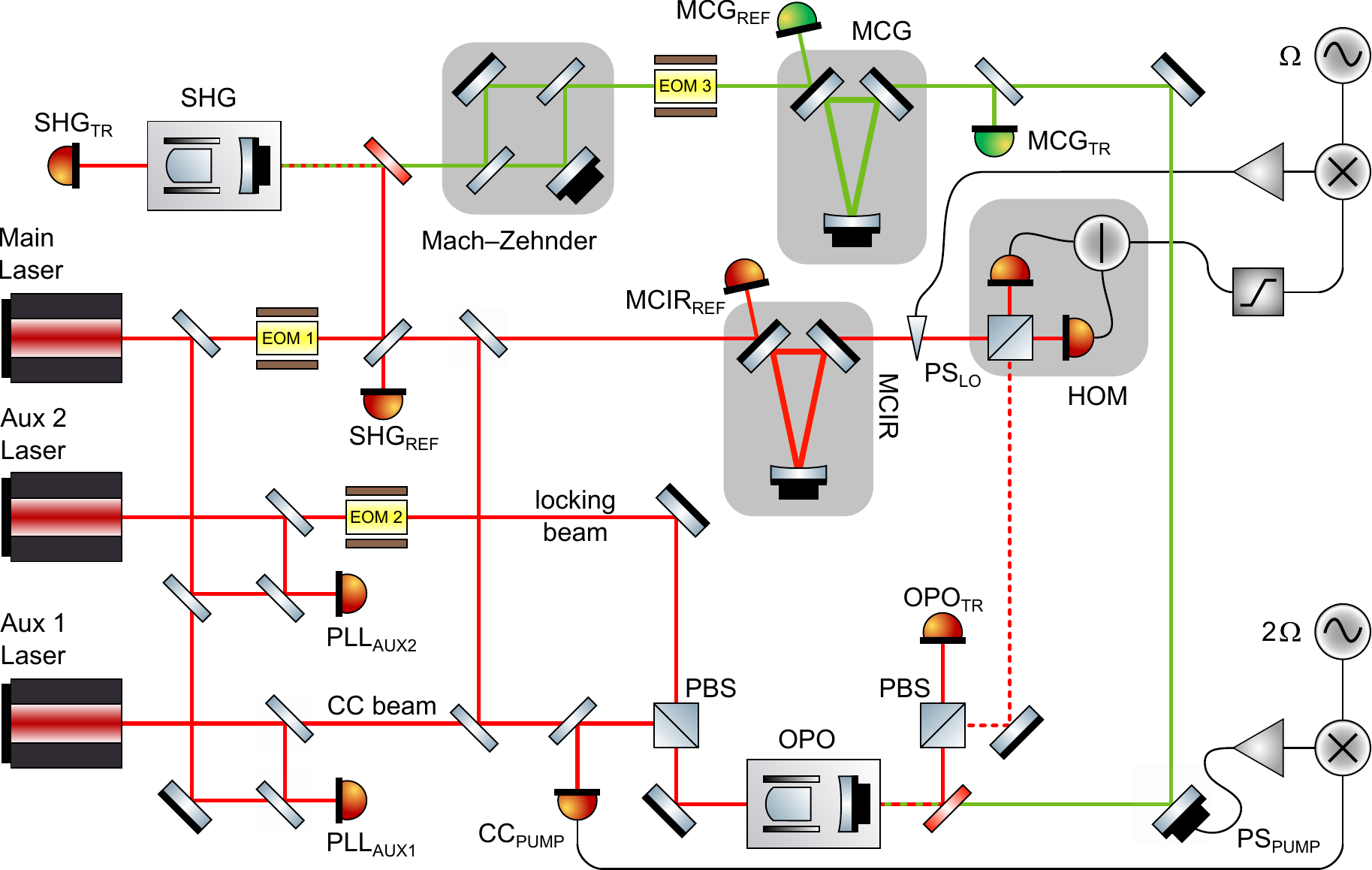}
  \end{minipage}\hfill
  \begin{minipage}[b]{0.29\textwidth}
    \caption{Optical layout of the current squeezing experiment controlled by the software. The optical layout is explained in section~\ref{sec:apparatus}. All photo-detectors used for the cavity locking and the \acs{CC} components are represented in the figure.} \label{fig:1500WoptLayout}
  \end{minipage}
\end{figure}

\noindent
The experimental setup (Fig.\ref{fig:1500WoptLayout}) is a \ac{FIS} source at the interferometer carrier wavelength i.e \SI{1064}{\nano\meter}, inspired by a similar system developed at GEO600~\cite{khalaidovski2012long}. In this setup, squeezed states of light are produced with an \ac{OPO} pumped with a second harmonic beam, depicted in green in Fig.\ref{fig:1500WoptLayout}. The pump beam is generated from a Main Laser in a \ac{SHG} cavity and it is intensity-stabilized by a \ac{MZ} interferometer. The generated squeezed states of light are measured on a \ac{HOM} using a strong coherent \ac{LO} beam. Higher order spatial modes of the pump and \ac{LO} beams are filtered using triangular cavities, \ac{MCG} and \ac{MCIR}, respectively. 
Two auxiliary lasers (Aux 1, Aux 2) are used to lock the \ac{OPO} cavity and to implement a \ac{CC} of the squeezing angle~\cite{PhysRevA.75.043814}.

\subsection{Controls requirements}
\noindent
This complex optical setup requires dedicated control electronics, with the advantage of a high flexibility given by software-driven hardware. For proper operation, auxiliary lasers are locked to the Main laser using two \acp{OPLL}.

Furthermore we stabilize actively the temperature of the \ac{OPO} and the \ac{SHG} crystals with a dedicated digital hardware control loops since the phase-matching in nonlinear crystals, used in such cavities, occurs only at given temperature.
Finally, we control the length of four optical resonators (\acs{SHG}, \acs{OPO}, \acs{MCIR}, \acs{MCG}), the power of the pump beam, the phase of the generated squeezed state, and the phase of the \ac{LO} at \ac{HOM}.
Thus, multiple control-loops should be designed to lock all loops in sequence in the shortest possible time for a high duty-cycle of operation. The system must provide the possibility of a parallel work on different parts of the experiment simultaneously. Furthermore a high degree of compatibility with the Virgo software environment is desirable, in view of potential future integration. 

\subsection{Controls design}
\begin{figure}[ht]
\begin{minipage}[t]{0.45\linewidth}
\centering
\includegraphics[width=\columnwidth]{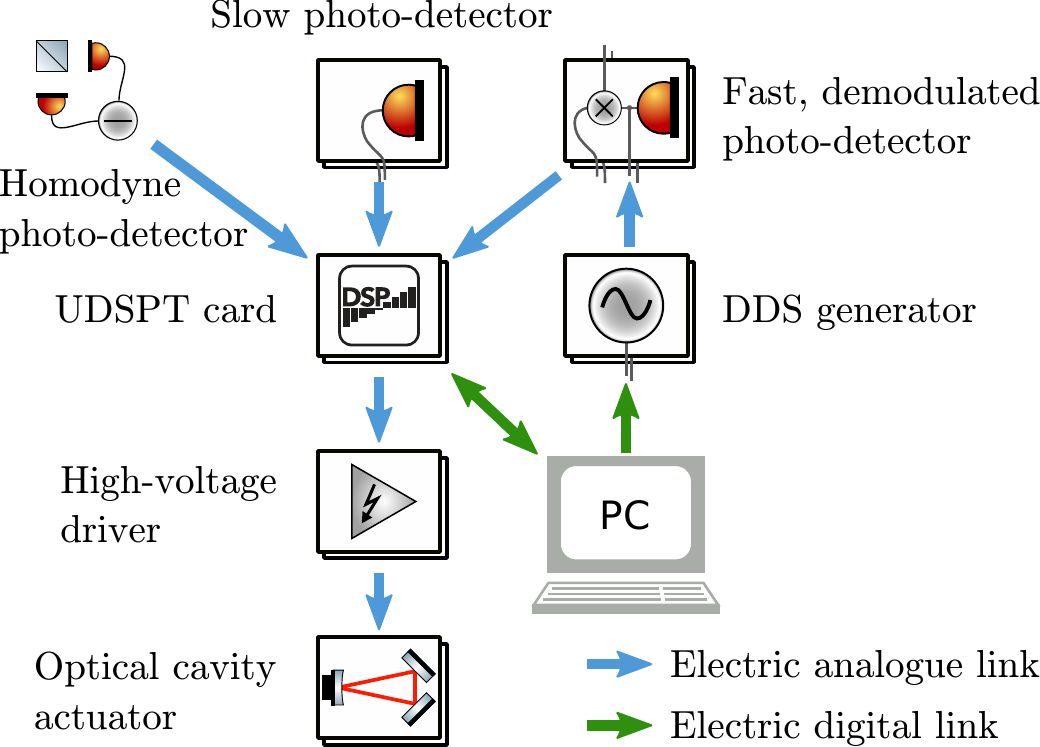}
    \caption{Hierarchy and information flow in the control management system of optical cavities. Numerous components are placed inside the pile symbol.}
    \label{fig:Electronic_system}
\end{minipage}
\hspace{0.3cm}
\begin{minipage}[t]{0.45\linewidth}
   \centering
   \includegraphics[width=\columnwidth]{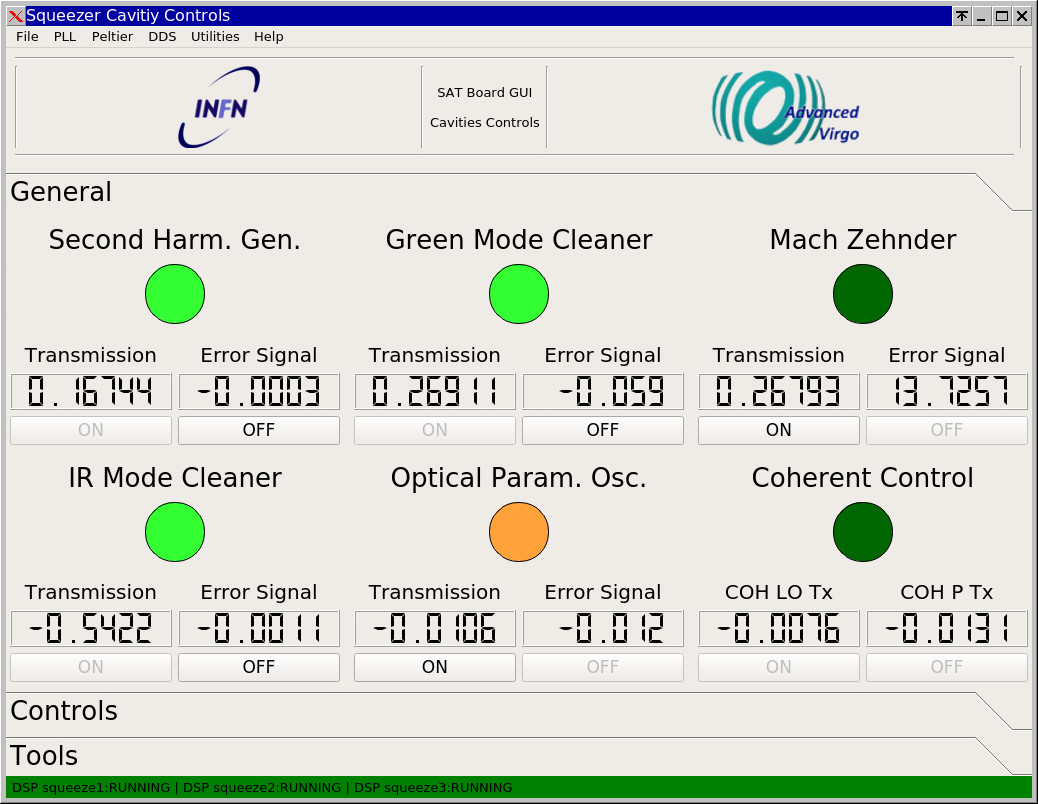}
   \caption{Screenshot of the sqzGUI interface for the interaction with the UDSPT software via Tango servers.}
   \label{fig:sqzGUI}
\end{minipage}
\end{figure}

\noindent
In our solution, \acp{OPLL} and temperature controls are locked by independent and sufficiently stable devices, so that additional automation is not required for these elements. On the other hand, the length of the four optical cavities (\acs{SHG}, \acs{OPO}, \acs{MCIR}, \acs{MCG}), the power of the green beam and both \acp{CC} loops are managed by the \ac{UDSPT}. The length control is implemented using \ac{PDH}~\cite{black2001introduction}, while the \ac{MZ} cavity, and both \ac{CC} loops are stabilized by a \ac{PID} controller. The structure of the connections to a \ac{UDSPT} hardware is shown in Fig.~\ref{fig:Electronic_system}. \acp{UDSPT} implements lock logic, calculates the feedback correction signals and drives piezo-electric actuators. Tunable \ac{DDS} generators provide necessary modulation and demodulation signals where needed. A dedicated GUI (Fig~\ref{fig:sqzGUI}) is designed to configure devices remotely. In the next paragraph, we describe the hardware which computes control algorithms.

\subsection{Hardware}
\label{sec:Hardware}
\noindent
The core component of the system are the \ac{UDSPT} cards~\cite{Virgo_second-generation}. They were developed for the real time control of the Virgo super attenuator suspension system. Each card has six \SI{24}{\bit} \acp{ADC} and six \acp{DAC} sampled at a maximum frequency of \SI{320}{\kilo\hertz}. The software of each card, developed in the \textit{damping-adv} \ac{SDK}, is executed in the time between sampling. An integrated \ac{DSP} unit implements efficient, high-order digital loop filters. \ac{UDSPT} cards also provide advanced connectivity between themselves, with the Virgo time distribution network and \ac{DAQ} system~\cite{DAQ_2008}. Here, we report the newest feature of the \ac{SDK} which implements the \ac{FSM} engine. We use \acp{FSM} for a quick lock acquisition of feedback loops. In order to manage locking loops, we use three cards: squeeze1, squeeze2 and squeeze3. squeeze1 distributes timing and exchanges information with \ac{DAQ}, squeeze2 manages devices related to the green light path and squeeze3 takes care of \ac{MCIR}, \ac{OPO} and \ac{CC}. The algorithm is described in Section~\ref{sec:Software}.

\section{Software description}
\label{sec:Software}
\noindent
The presented software is composed of two parts: a Python-based \ac{GUI}, called sqzGUI, and a software executed by the \acp{UDSPT}, called DSPcode--Squeezing. The two pieces communicate via Tango Controls server~\cite{Tango2}, allowing the simultaneous run of many instances of sqzGUI. Each instance has an equal priority and enables a synchronous co-working. The \ac{GUI} appearance is shown in Fig.~\ref{fig:sqzGUI}. The sqzGUI integrates the control of every device involved in the experiment. The main loop is executed in one second interval. It implements the slow part of the management logic assigned to the \acp{UDSPT}, driving namely: \ac{MCIR}, \ac{OPO}, \ac{SHG}, \ac{MCG}, \ac{MZ}, the \ac{CCpump} and the \ac{CClo} loops.

\begin{figure}[ht]
\centering
\includegraphics[width=\columnwidth]{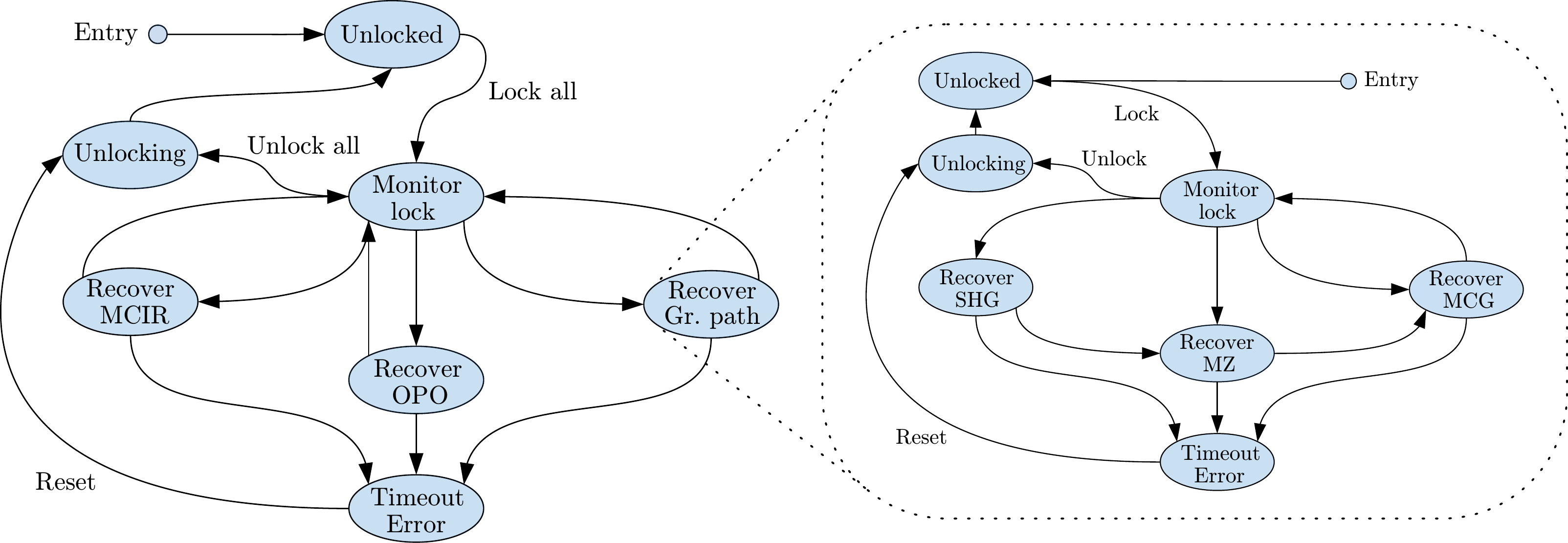}
   \caption{\acs{FSM} for the lock monitoring and the recovery of the whole bench. When the machine is launched, it enters the \textit{Unlocked} state. When in \textit{Monitor lock}, states are switched automatically. There are three manually triggered transitions: \textit{Lock all}, \textit{Unlock all} and \textit{Reset}. The machine leaves \textit{Monitor lock} automatically when any of the monitored devices need to be recovered. Another \acs{FSM} is implemented in \textit{Recover Gr. path} for the lock monitoring and the recovery of the green light path. In this machine, the lock sequence is maintained in order to ensure a correct lock of subsequent devices.}
   \label{fig:FSM_Main}
\end{figure}

Communication of sqzGUI with each \acp{FSM} in the DSPcode occurs via a couple of unidirectional variables: BUTTON and STATUS. The functions for the communication are gathered in the DSP\_sq module. The structure of the slow automation logic is split in two and shown in Fig.~\ref{fig:FSM_Main}. \ac{FSM} supervising the green light path is nested in the main \ac{FSM} and it supervises the lock of the three devices in series conditioning green light.

\begin{figure}[ht]
    \centering
   \includegraphics[width=\columnwidth]{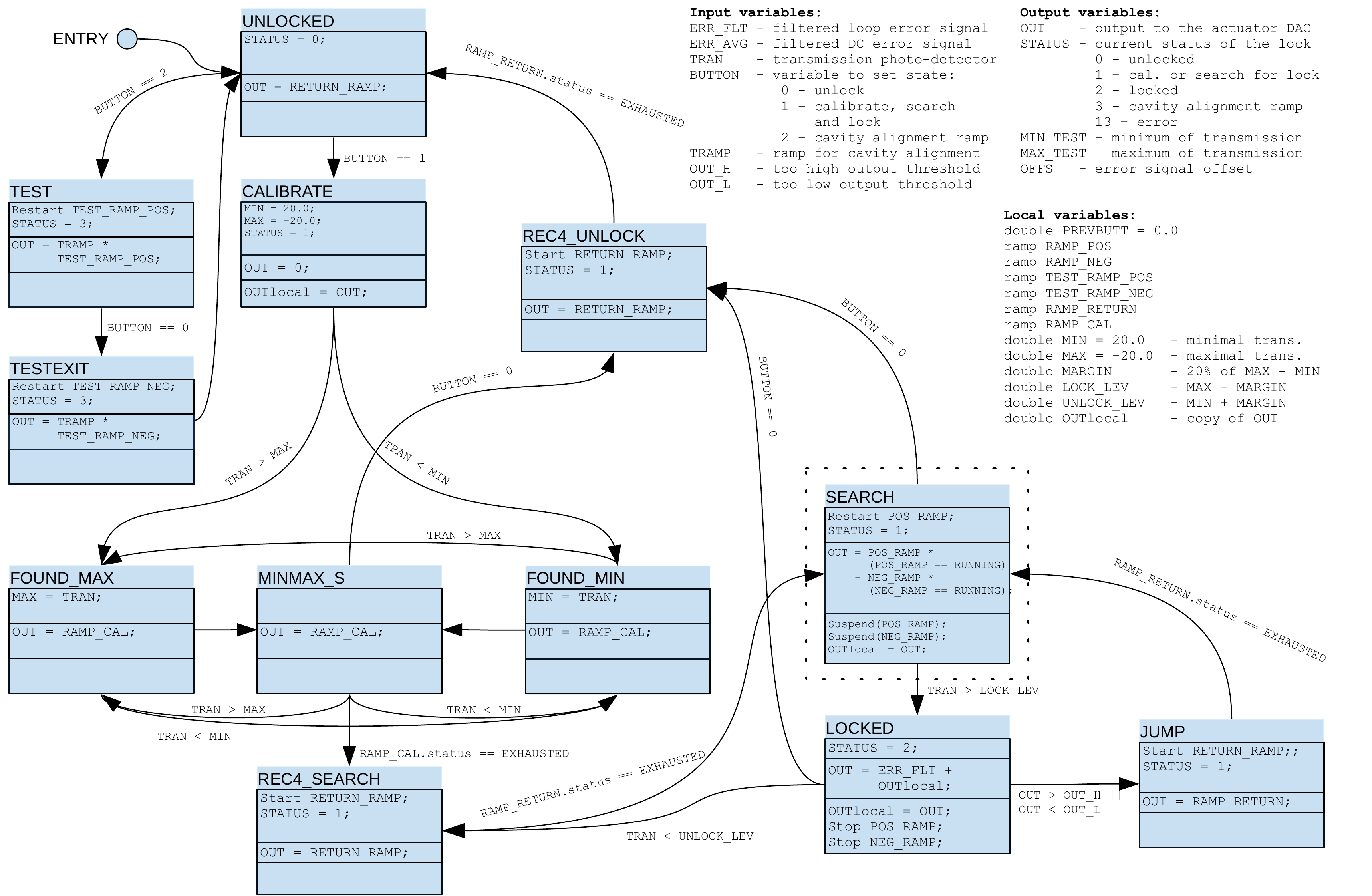}
    \caption{Generic block diagram of the \acs{FSM} used for fast lock acquisition, monitoring and automatic recovery of a single optical loop.}
    \label{fig:FSM_4cavity}
\end{figure}

The sqzGUI python code is self-documented according to PEP~257. Since \textit{damping-adv} does not support self documentation we focus on the \ac{UDSPT} code in more details. Nevertheless, comments are included in the code in the most important sections. Each optical cavity, the \ac{MZ} interferometer and the \ac{CC} loops are managed by a dedicated \ac{FSM} implemented in the \acp{UDSPT}. The structure of this \ac{FSM} is shown in Fig.~\ref{fig:FSM_4cavity}.

In this graph, the SEARCH state is needed only in case of a resonant cavity lock, hence it is circled with a dashed line. In case of a \acs{PID} lock, the SEARCH state does not exist and transitions are straightforward. In the figure, the state of the machine is represented as a rectangle divided in three sections with a name of the state above. The first section contains the instructions performed at entry to the state, the middle section encloses the instructions executed every time the \acs{FSM} algorithm is calculated and the machine is in this state, the last section describes the actions taken when leaving the state. Transitions between states are depicted with arrows, accompanied by the condition of the transition written near the arrow. The \acs{FSM} can exchange signals with the outer world via one-directional variables declared as input or output. Instead, local variables are exclusive for a particular instance of the machine and their value is preserved between states. A particular type of a variable is \textit{ramp} which simplifies the generation of ramps. \textit{ramp} is an object which stores the current value of the ramp, the status of the ramp and it implements three methods: \textit{Start}, \textit{Stop} and \textit{Restart}.
The \ac{FSM} algorithm is executed in the time between consecutive samples of the ADC. Thus, its complexity may limit the maximal sampling frequency. In our case, the \acp{ADC} sample at \SI{160}{\kilo\hertz}.

\begin{figure}[ht]
\centering
\includegraphics[width=0.8\columnwidth]{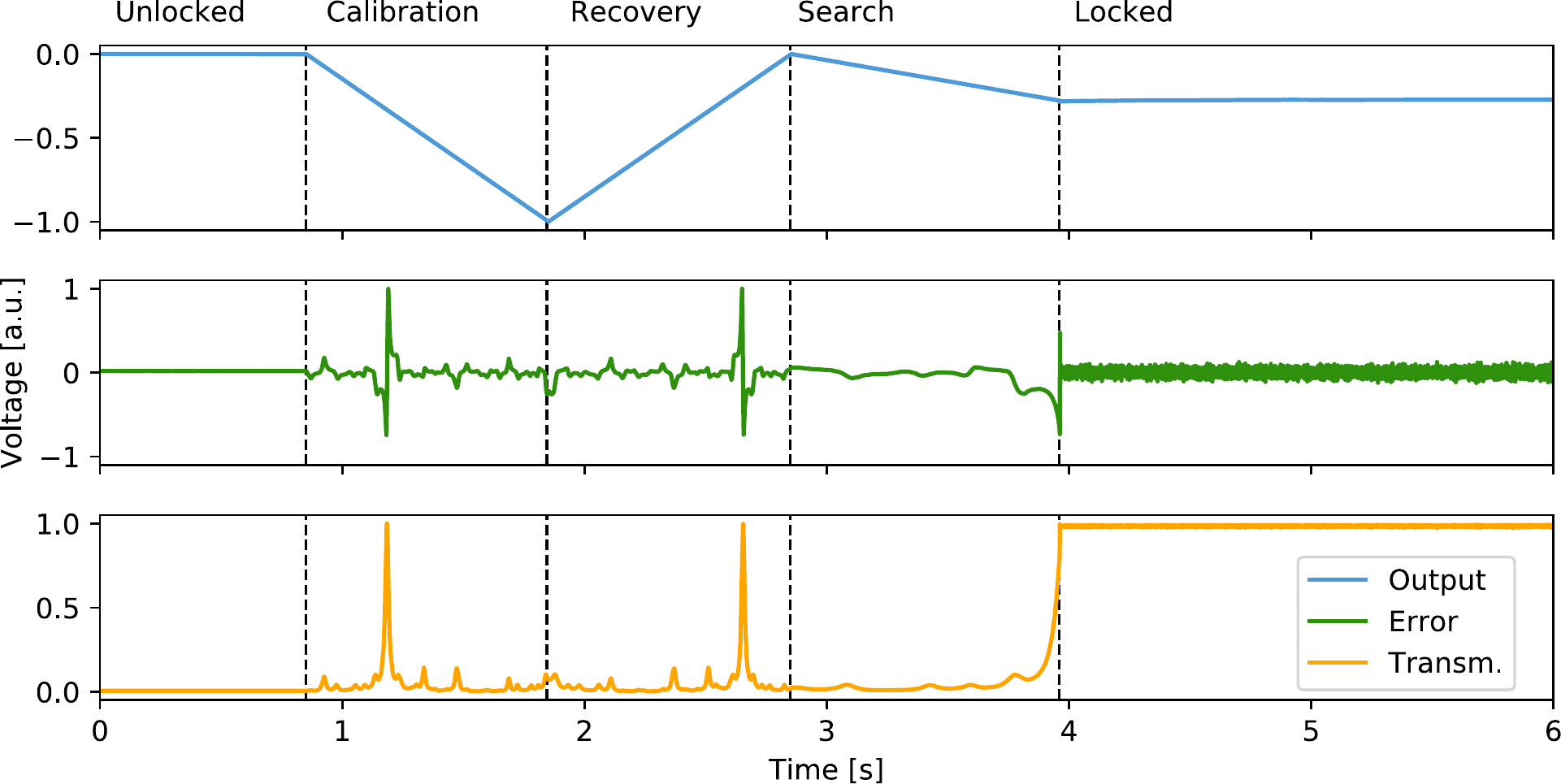}
   \caption{Locking sequence of an optical cavity.}
   \label{fig:Lock_sequence}
\end{figure}

Now, we look at the logic of the lock. At the beginning of the algorithm, the \ac{FSM} starts in an UNLOCKED state. It is an idle state where the output is zero. Some transitions occur only if the BUTTON value changes. From the UNLOCKED state, one of the two different paths is enabled; the generation of a ramp at the output for a cavity alignment (TEST) or an attempt of lock. In the former case, we set the BUTTON value to ``2''. The output is currently a product of the internal ramp generator and the ramp slope TEST\_RAMP\_POS going from 0 to 1, assuring smooth rise of the signal at the piezo-actuator. At the end of the alignment, we set BUTTON to ``0'' and the output fades out as TEST\_RAMP\_NEG varies from 1 to 0. The transition to locking is only possible from UNLOCKED. Setting BUTTON to ``1'', the machine goes first through a calibration process. To recognize the proper locking threshold, indicated by the cavity transmission, the machine scans the output and commutes between MINMAX\_S, FOUND\_MAX and FOUND\_MIN registering the highest and the lowest level of transmission. After the calibration, we assume that \SI{20}{\percent} of the minimum and maximum are the right levels for unlock and lock detection correspondingly. In REC4\_SEARCH, the output of the machine turns back to zero to enable SEARCH. The machine remains in the SEARCH state, generating a ramp signal at the output until the locking level is not reached. Search can be interrupted by a manual command. When the LOCKED state is reached, the machine substitutes the ramp at the output with the digitally filtered input signal. The input signal is provided from the demodulated output of the fast photo-detector. There are three possible exits from the LOCKED state: a manual interruption to UNLOCKED, a lock loss when the transmission drops below the unlock level to SEARCH and a technical detail of the implementation when the loop purposely unlocks and moves the output back to the center of the dynamics of the piezo driver via JUMP state. To re-lock after the jump, the machine passes through SEARCH. An example of locking sequence is shown in Fig.~\ref{fig:Lock_sequence}.

For the visualisation of digitized signals in the experiment, we use dataDisplay software~\cite{dataDisplay}. This tool is developed for the Virgo experiment. It intends to process and visualize the online data with low latency, to facilitate the commissioning and the analysis of the Virgo interferometer.

\section{Conclusions}
\noindent
We presented the first application of the new feature of \textit{damping-adv} \ac{SDK}. The software significantly speeds up the work on the optical bench thanks to its multi-user architecture, automation and remote access. The complete optical bench for squeezed states of light generation has been automatized and it is ready for future research and development of improved squeezing techniques for gravitational-wave detectors. Thanks to the software-based approach, the existing control system can be easily scaled for large size optical experiments such as, for the alternative use of the squeezed states in gravitational-wave detectors~\cite{proposalEPR, Valeria, Sibilla}.

\begin{acronym}[MCIR]
\acro{RF}{Radio Frequency}
\acro{AF}{Audio Frequency}
\acro{PC}{Personal Computer}
\acro{FSM}{Finite-State Machine}
\acro{GUI}{Graphical User Interface}
\acro{PLL}{Phase-Locked Loop}
\acro{OPLL}{Optical Phase-Locked Loop}
\acro{DSP}{Digital Signal Processing}
\acro{DDS}{Direct Digital Synthesis}
\acro{PDH}[PDH technique]{Pound--Drever--Hall technique}
\acro{TEC}{Thermoelectric Cooler}
\acro{PID}{Proportional--Integral--Derivative} 
\acro{IT}{Information Technology}
\acro{QoS}{Quality of Service}
\acro{VCO}{Voltage-Controlled Oscillator}
\acro{DAQ}{Data AcQuisition}
\acro{CMRR}{Common-Mode Rejection Ratio}
\acro{SDK}{Software Development Kit}
\acro{ADC}{Analog-to-Digital Converter}
\acro{DAC}{Digital-to-Analog Converter}
\acro{RBW}{Resolution Bandwidth}
\acro{DC}{Direct Current}
\acro{AC}{Alternating Current}
\acro{CCpump}{pump beam coherent control}
\acro{CClo}{local oscillator coherent control}
\acro{EOM}{Electro-Optic Modulator}
\acro{UDSPT}{Digital Signal Processing unit}
\acro{HV}{High Voltage}

\acro{IR}{Infra Red}
\acro{HR}{High Reflectivity}
\acro{QN}{Quantum Noise}
\acro{GW}{Gravitational Wave}
\acro{SN}{Shot Noise}
\acro{RPN}{Radiation Pressure Noise}
\acro{SQL}{Standard Quantum Limit}
\acro{BAB}{Bright Alignment Beam}
\acro{LO}{Local Oscillator}
\acro{CC}{Coherent Control}
\acro{FSR}{Free Spectral Range}
\acro{FWHM}{Full Width at Half Maximum}
\acro{OPO}{Optical Parametric Oscillator}
\acro{OPA}{Optical Parametric Amplification}
\acro{SHG}{Second-Harmonic Generation}
\acro{MCIR}{Infrared Modecleaner}
\acro{MCG}{Green Modecleaner}
\acro{MZ}{Mach--Zehnder}
\acro{PPKTP}{Periodically Poled Potassium Titanyl Phosphate}
\acro{RoC}{Radius of Curvature}
\acro{TEM}{Transverse-Electo-Magnetic}
\acro{CCB}{Coherent Control Beam}
\acro{HOM}{Homodyne detector}
\acro{FIS}{FrequencyIndependent Squeezing}
\end{acronym}  



\end{document}